\documentclass[preprint,bibnotes,showpacs,preprintnumbers,amsmath,amssymb,prd,superscriptaddress]{revtex4}  

\usepackage{url}
\usepackage{graphics}
\usepackage{dcolumn}
\usepackage{bm}

\begin{document}
%

\title{Big-Bang Nucleosynthesis in comparison with observed helium and deuterium abundances -- possibility of a non-standard model}

\author{R. Ichimasa
\footnote{E-mail: ichimasa@phys.kyushu-u.ac.jp}}
\affiliation{Department of Physics, Kyushu University, Hakozaki,
Fukuoka 812-8581, Japan}

\author{R. Nakamura}
\affiliation{Department of Physics, Kyushu University, Hakozaki,
Fukuoka 812-8581, Japan}
\affiliation{
Kurume Institute of Technology, Kamitsu-machi, Kurume, 830-0052,
Japan}

\author{M. Hashimoto}
\affiliation{Department of Physics, Kyushu University, Hakozaki,
Fukuoka 812-8581, Japan}

\author{K. Arai}
\affiliation{
Department of Physics, Kumamoto University, Kurokami,
Kumamoto 860-8555, Japan
}

%



\begin{abstract}
Comparing the latest observed abundances of $^4$He and D, we make a $\chi^2$ analysis
to see whether it is possible to reconcile primordial nucleosynthesis using up-to-date 
nuclear data of NACRE~II and the mean-life of neutrons. 
If we adopt the observational data
of $^4$He by Izotov et al.~\cite{Izotov2013}, 
we find that it is impossible
 to get reasonable concordance
against the standard Big-Bang nucleosynthesis.
However, including degenerate neutrinos, we have succeeded in
obtaining consistent constraints between the neutrino degeneracy and the
baryon-to-photon ratio from detailed comparison of calculated abundances with
the observational data of $^4$He and D: the baryon-to-photon ratio in
 units of $10^{-10}_{}$
is found to be in the range $6.02 \lesssim \eta_{10} \lesssim 6.54$ 
for the specified parameters of neutrino degeneracy.

\end{abstract}

\date{\today}

\maketitle

\section{Introduction}
Big-bang nucleosynthesis (BBN) provides substantial clues for
investigating physical conditions in the early universe.
Standard BBN produces about 25 \% of mass in a form  of $^4$He, which
has been considered to be in good agreement with its abundance  observed
in a variety of astronomical objects~\cite{Iocco:2008va,Steigman2007,Coc2012,Coc2013}.
The produced amount of $^4_{}$He depends strongly on a fraction of
neutrons at the onset of nucleosynthesis, but is not 
very sensitive to 
the baryon-to-photon
 $\eta$~($\eta = n^{}_{\rm b}/n^{}_\gamma;\, \eta^{}_{10} = 10^{10}_{} \eta$). 
Hence the produced amount of $^4$He is used to explore the expansion rate during BBN, which can be
related to the effective number of neutrino flavours~\cite{Simha2008}.
In addition to  ${}^4_{}$He, significant amounts of D, ${}^3_{}$He and ${}^7_{}$Li are also produced. Because of its strong dependence on $\eta$, the abundance of D is crucial in determining $\eta$ and consequently the density parameter of baryons $\Omega^{}_b$.

In spite of apparent success in standard BBN, 
recent observed light elements considered to be primordial
have been controversial.
Large discrepancies for $^4$He observations emerge between different observers and modelers of observations:
Rather high values of $^4$He have been reported for H II regions in blue
compact galaxies~\cite{Aver2012,Izotov2013}.
It is noted that primordial abundance of $^4$He is deduced from
extrapolation to the zero metallicity~\cite{Aver2013}.
 Deuterium abundance has been observed in absorption systems toward high
 redshift quasars~\cite{Cooke2014}.
It should be noted that the value in D  has been believed
to limit the present baryon density (e.g. Schramm \& Turner~\cite{schr}).
A low value of ${}^7_{}$Li observed in Population II stars reported by Bonifacio et al.~\cite{Bonifacio2007}
is considered to be due to depletion and/or destruction during the lifetimes of stars
from a high primordial value~\cite{Korn2006,Melendez2010}.

{{{Recently, the half-life of neutrons has been updated from the previous adopted value of 
$885.7 \pm 0.8$~s~\cite{PDG2008} which has been used commonly in BBN calculations consistent with the observed abundances of ${}^4$He and D.}}}
However, the latest compilation by 
Beringer et al. derives the
mean-life to be $880.1 \pm 1.1$~s~\cite{PDG2012}, 
which may suggest inconsistency between BBN and observational values.
This indicates further inconsistency against $\eta$ deduced by
\cite{Planck_basic,Planck_cosmo}.

The apparent spread in the observed abundances of $^4$He should give rise to an inconsistent range of $\eta$. 
Apart from observational uncertainties, we have no reliable theories beyond the standard theory of elementary particle physics. 
It is assumed in standard BBN that there are three flavours of massless neutrinos which are not degenerate. However it is suggested by Harvey and Kolb~\cite{Harvey1981} that lepton asymmetry could be large even when baryon asymmetry is small. The magnitude of the lepton asymmetry is of particular interest in cosmology and particle physics. Related to neutrino oscillations, investigations of BBN have been 
reprised with use of non-standard models (e.g.~Ref.~\cite{sark_b}).
As presented by Wagoner et al.~\cite{wago67} and Beaudet \& Goret~\cite{beau76}, 
the abundances of light elements are modified by neutrino degeneracy
(see previous investigations~\cite{yang84,Terasawa85,Kang92,Lisi99}, see also review by Ref.~\cite{sark_a}); 
it could be necessary and crucial to search consistent regions
 in $\eta$ within a framework of BBN with degenerate neutrinos by
comparing with the latest observation of abundances of He and D.

If neutrinos are degenerate, the excess density of neutrinos causes
speedup in the expansion of the universe, leaving more neutrons and
eventually leading to enhanced production of $^4$He. On the other hand,
degenerate electron-neutrinos shift $\beta$-equilibrium to less or more
neutrons and hence change abundance production of $^4$He. The latter effect is more
significant than the former.
In the present paper we investigate BBN with including degenerate
neutrinos and using up-to-date nuclear data. Referring to several sets
of combinations
for recent observed abundances of $^4$He and D, we derive consistent 
constraints between $\eta$ and the degeneracy parameter. 

In \S~\ref{sec:obs} we summarize the current situation of observed
abundances of light elements. Our results of BBN with 
updated nuclear data are presented in \S~\ref{sec:BBN}. 
Discussion is given in \S~\ref{sec:result}.

\section{Observed abundances of $^4$He, D, and $^7$Li}
\label{sec:obs}

There exist very large spreads in some observed abundances of light elements
 due to  different observational methods.
Let us describe how we adopt the 
observed primordial {abundances}.


The primordial abundance of ${}^4$He can be measured from
observations of the helium and hydrogen emission lines from low
metallicity blue compact dwarf galaxes. Izotov et al.
reported the ${}^4$He abundance from a subsample of 111 HII regions as follows~\cite{Izotov2013}:
\begin{equation}
 Y^{}_p = 0.254 \pm 0.003. \label{eq:Yp_Izotov13}
\end{equation}
It should be noted that primordial abundance of $^4$He could be appreciated to the zero-metalicitiy
in terms of an extrapolation by a model of chemical evolution of galaxies.
An alternative low value on the average is reported by Aver et al.~\cite{Aver2013}:
\begin{equation}
 Y_p = 0.2464 \pm 0.0097 \label{eq:Yp_Aver13}
\end{equation}
which has a very large spread in errors.

Deuterium is the most crucial element to determine $\eta$ because of the
strong and monotonic dependence on $\eta$.
Its primordial abundance is determined from metal-poor absorption
systems toward high redshift quasars. Cooke et al. have performed measurements
at redshift $z = 3.06726$ toward QSO SDSS J1358$+$6522~\cite{Cooke2014}. 
Additionally, they have analysed all of the known deuterium
absorption-line system that satisfy a set of strict criteria,
\begin{equation}
 {\rm D/H} = \left( 2.53 \pm 0.04 \right) \times 10^{-5}.
\label{eq:D_Cooke14}
\end{equation}
This value corresponds to the baryon density $\Omega^{}_bh^2=0.02202\pm0.00046$
which is consistent 
with the results of Planck experiment~\cite{Planck_basic,Planck_cosmo}.
Here $h$ is the Hubble constant in units of $100$~km/s/Mpc.

We should note that the observed abundance of $^7$Li in Population II
stars is given by Sbordone et al. to be~\cite{sbord}:
\begin{equation}
 {\rm {}^{7}Li/H} = \left( 1.58 \pm 0.31 \right) \times 10^{-10}, 
\label{eq:Li7_Sbord}												
\end{equation}
which has been advocated to be rather low compared with  BBN.
While, considering significant depletion and/or destruction during the
lifetimes of Population II stars, 
Korn et al. have derived a high  primordial abundance~\cite{Korn2006}:
\begin{equation}
{}^7{\rm Li/H} = \left( 2.75 - 4.17 \right)\times10^{-10}, 
\label{eq:Li7_Korn}
\end{equation}
{{a value which}}  is still too low to reconcile with the result of BBN.
It is noted that Li can be produced together with Be and B through
spallation  of CNO nuclei by cosmic ray protons and
$\alpha$-particles. About 10 \% of $^7$Li could be due to cosmic ray
processes leaving remainder as primordial~\cite{Mene71,ol_b}.  
 Among a variety of observational data, we here pick up only
 representatives of ${}^4$He and D which we adopt in terms of symbols.

\section{Big-Bang nucleosynthesis}
\label{sec:BBN}

\begin{figure}[ht]
\resizebox{\hsize}{!}{\includegraphics{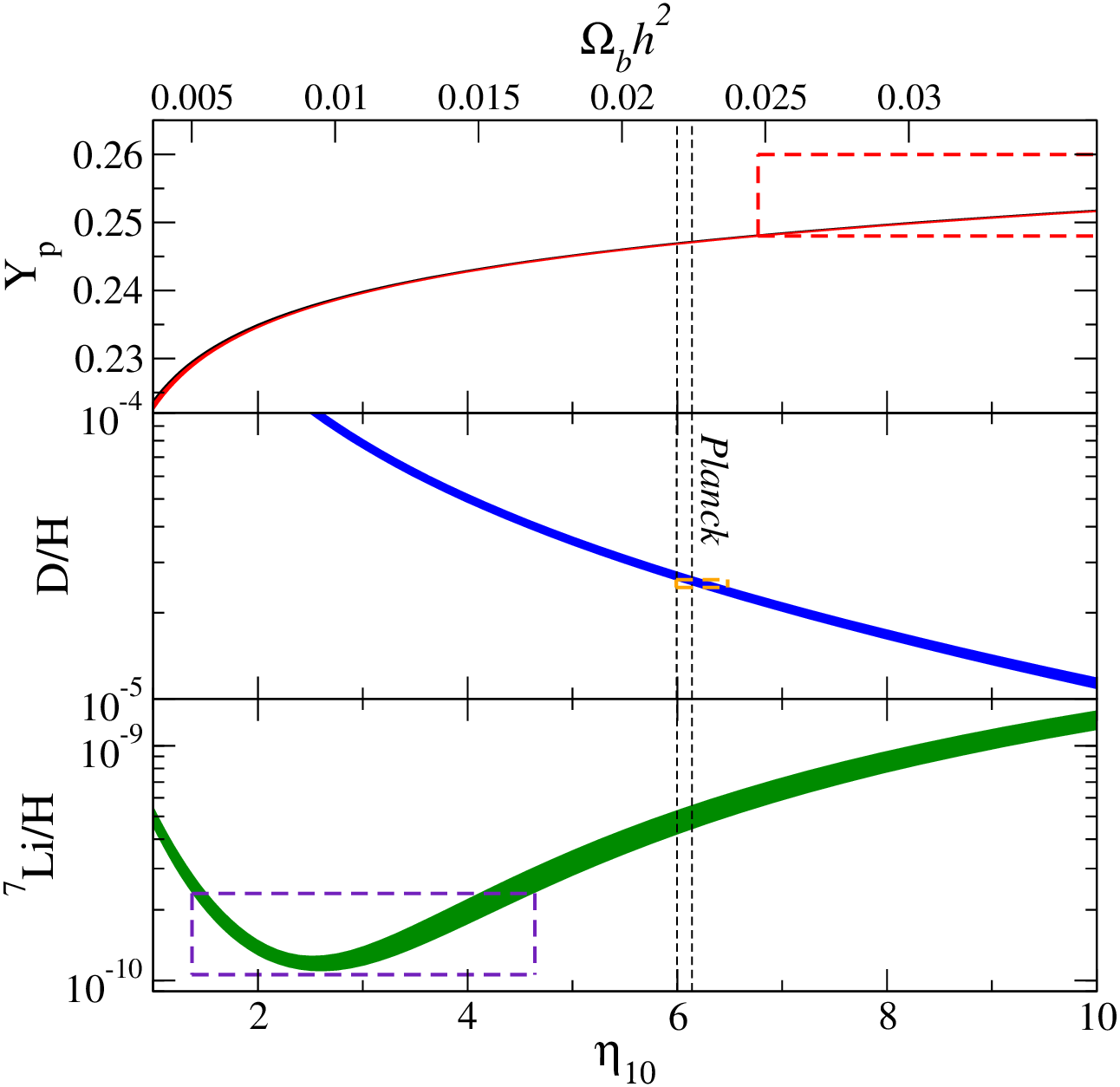}}
 \caption{Primordial abundances produced in a standard model as a function of 
$\eta_{10}$ with use of the nuclear data of NACRE~II and the mean-life of 
neutrons by Beringer et al.\cite{PDG2012}. The vertical band indicates the
 result of Planck~\cite{Planck_cosmo}. The boxes show the
 observational abundances of ${}^4_{}$He~\cite{Izotov2013},
 D/H~\cite{Cooke2014}, and ${}^7_{}$Li/H~\cite{sbord} with $2\sigma$ uncertainties.}
\label{fig:BBN_result_nacre2}
\end{figure}

\begin{figure}[th]
\resizebox{\hsize}{!}{\includegraphics{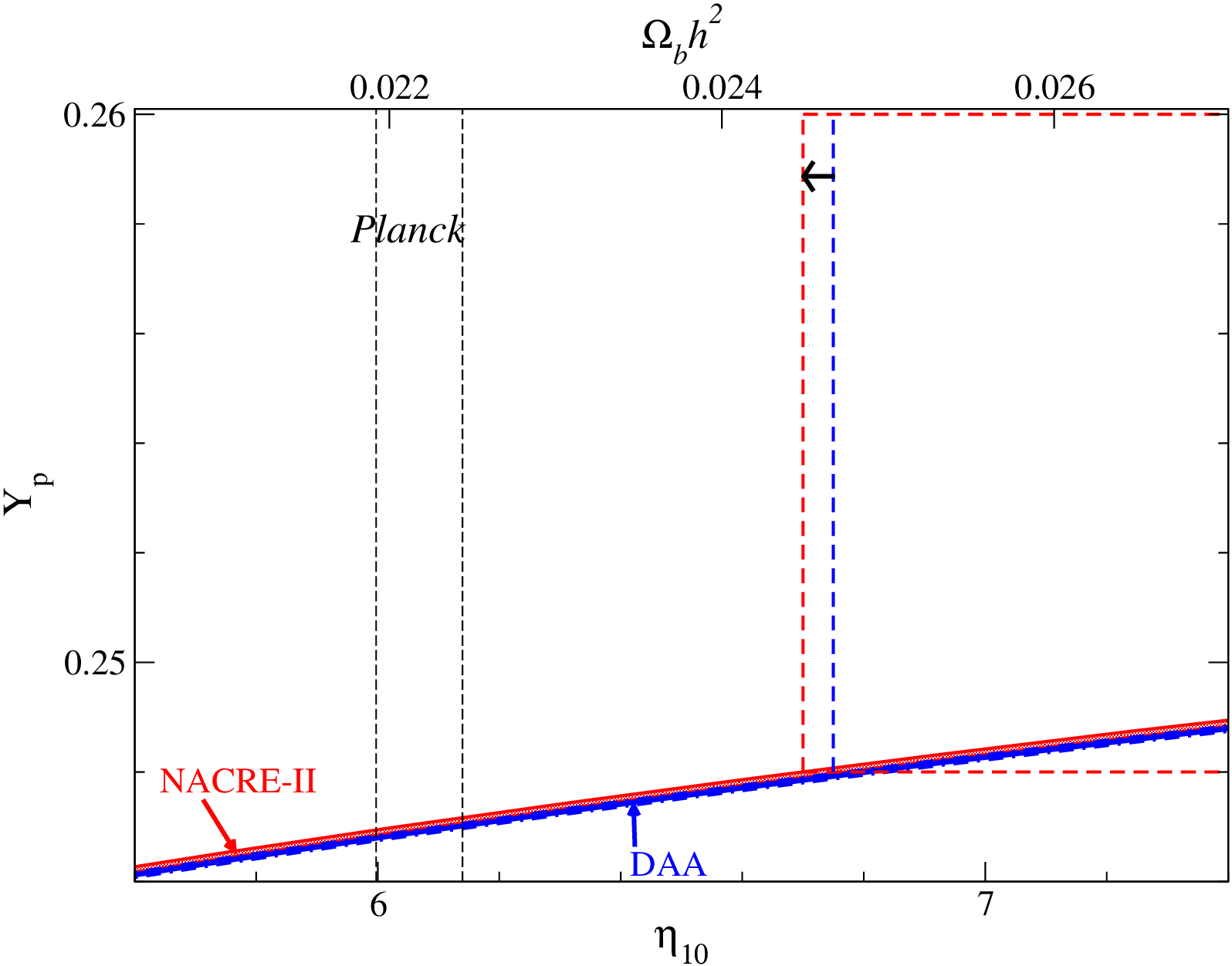}}
 \caption{Effects of reaction rates on the production of $^4$He. The red line is drawn 
using NACRE~II~\cite{NACRE2} and the blue one is due to DAA~\cite{DAA04}}
\label{fig:Yp_result_nacre2} 
\end{figure}

\subsection{Standard Big-Bang Nucleosynthesis}

Let us compare the calculated abundances in BBN with the observed ones. 
It is emphasized 
that standard BBN fails to find consistent range of $\eta$ for the observed values
{{given in}} \eqref{eq:Yp_Izotov13} and \eqref{eq:D_Cooke14} as explained below. 
Nucleosynthesis is calculated with use of a network constructed by Hashimoto 
\& Arai~\cite{Hashimoto1986}, where the reaction rates are taken from NACRE~II~\cite{NACRE2} supplemented by 
Descouvemont et al.(DAA)~\cite{DAA04}, Caughlan \& Fowler~\cite{CF88}, and
Ando et al.~\cite{Ando2006}. 

The mean-life of neutrons is taken to be $880.1$~s~\cite{PDG2012}.
Now the mean-life becomes drastically short compared to the previous value of 
$885.7$~s~\cite{PDG2008}. 
Using $\chi^2_{}$-analysis for measured mean-lives,
Beringer et al.~\cite{PDG2012} have obtained the up-dated (recommended) value
to be  $\tau_{\rm n}=880.1 \pm 1.1$~s within the $1\sigma$ level.

We set  the number of neutrino species to be 3 for simplicity.
We adopt the present CMB temperature of $2.725$~K~\cite{COBE1999}.

{{
In Fig.~\ref{fig:BBN_result_nacre2}, we compare observed abundances of
${}^4_{}$He, D, and ${}^7_{}$Li with BBN, assuming $1\sigma$ errors for the nuclear reaction rates.}}
We cannot find an overlapped region for the observational data between
He by Izotov et al.~\cite{Izotov2013} and D by Cooke et al.~\cite{Cooke2014}.
We also compare the baryon-to-photon ratio obtained
from our calculations with the range $5.98 \le \eta^{}_{10} \le 6.16$ derived
from {\it Planck} observation. 
Contrary to the concordance with Planck result for D, the abundances ${}^4$He and ${}^7_{}$Li
give no consistent range of $\eta$.

Figure~\ref{fig:Yp_result_nacre2} shows the uncertainties 
in the produced abundance of ${}^4_{}$He 
due to the alternative reaction rates of NACRE~II~ and DAA. 
{{
The difference $\delta\eta^{}_{10}\sim0.07$ between the two groups is very small and therefore
does not resolve the inconsistency.}}

\subsection{BBN with  neutrino degeneracy}


\begin{figure}[th]
\resizebox{\hsize}{!}{\includegraphics{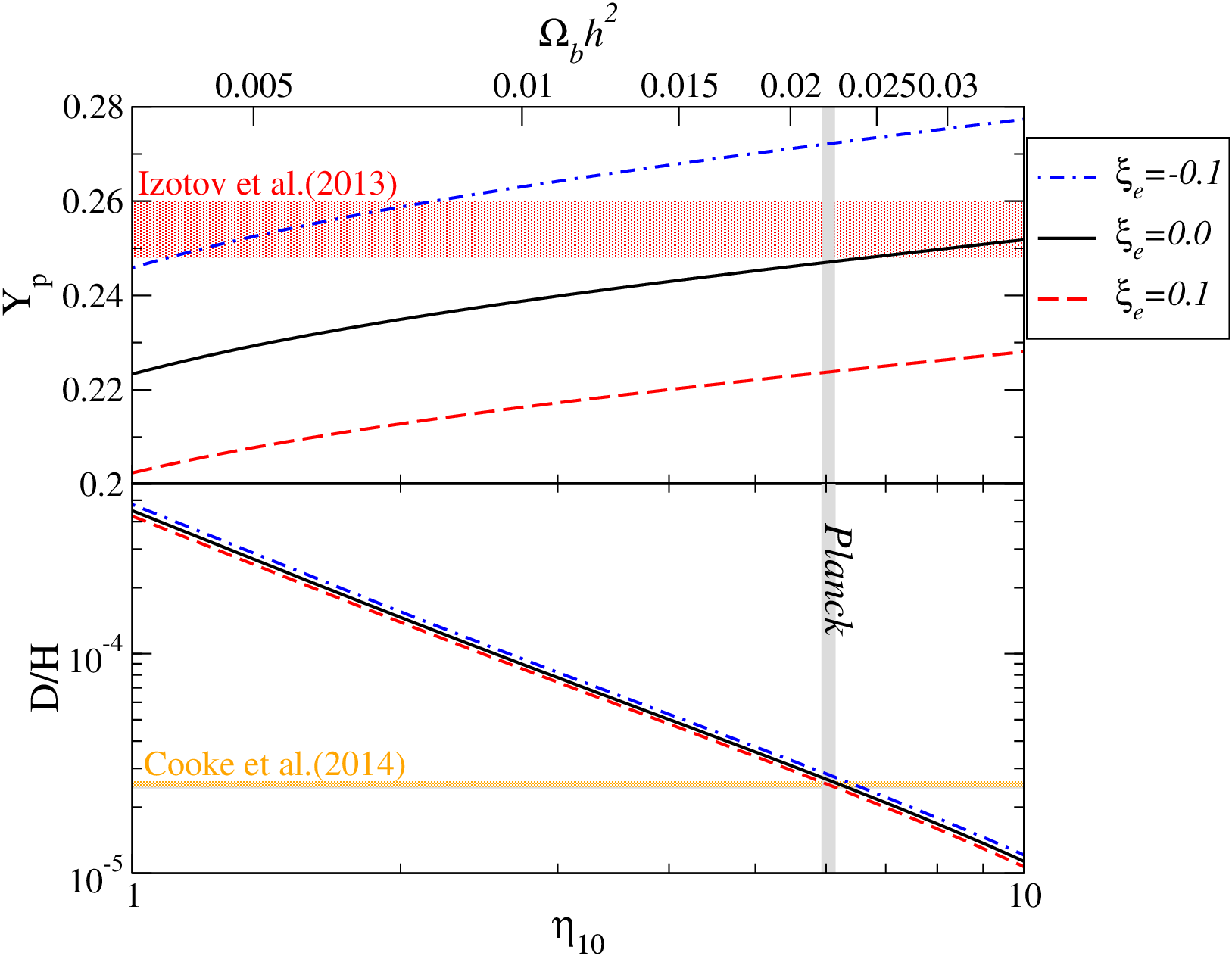}}
 \caption{Effects of neutrino degeneracy on the production of $^4$He and D/H. 
The degeneracy parameters is taken to be 
$\xi_{\rm e}=-0.1, 0$, and $0.1$ from the top to bottom
 curve. The vertical  band comes from the baryon density determined by $Planck$.
 The horizontal bands correspond to the observational abundances of
 ${}^4_{}$He and D/H with $2\sigma$ uncertainty. }
\label{fig:bbn_xi_nacre2}
\end{figure}

\begin{figure}[th]
\resizebox{\hsize}{!}{\includegraphics{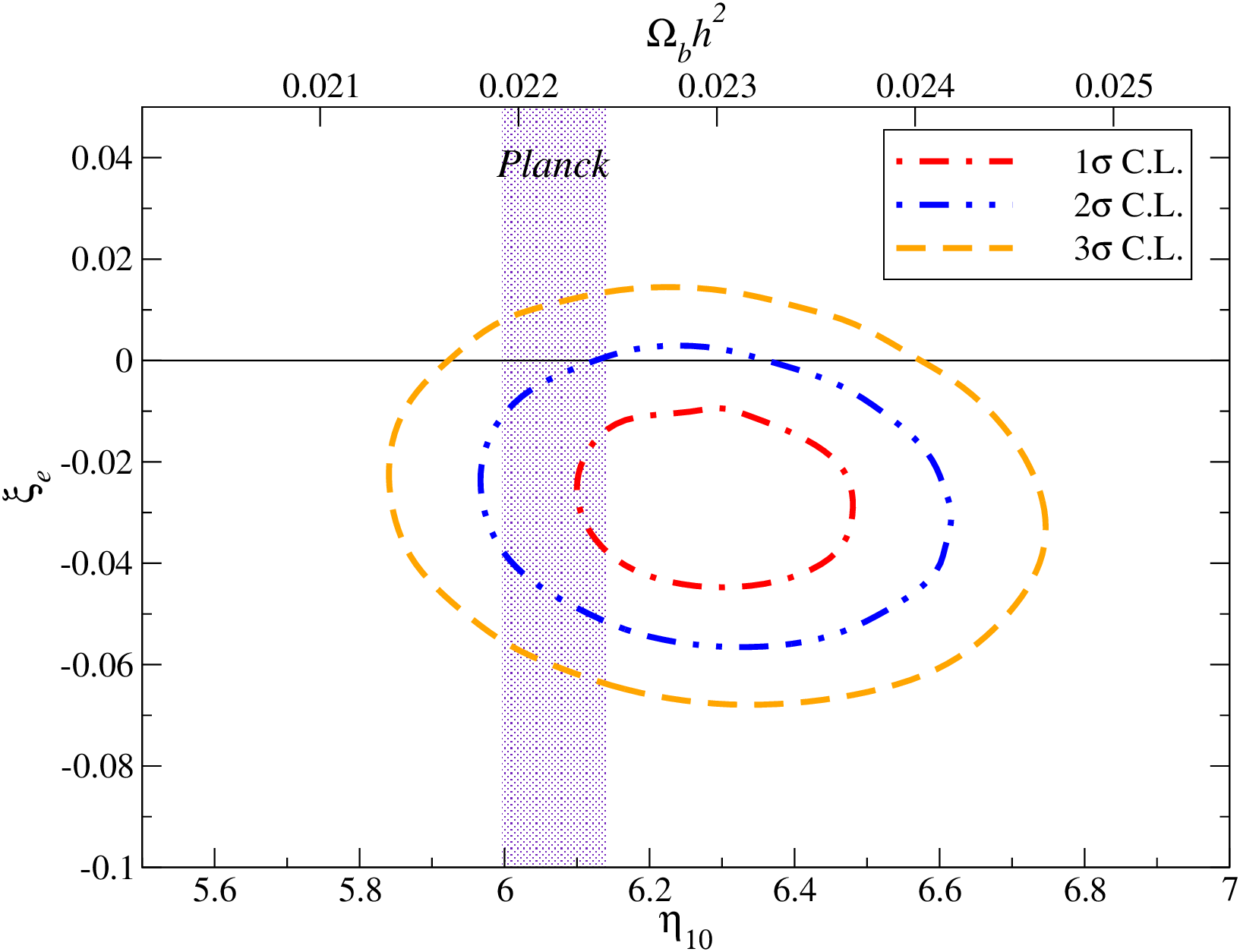}}
 \caption{Contours having 1$\sigma$, 2$\sigma$, and 3$\sigma$ confidence
 levels from $Y_p$ and D/H observations on the $\eta^{}_{10}-\xi_{\rm e}$ plane.
The horizontal line corresponds to SBBN~($\xi^{}_{e}=0$). The vertical
 band shows the baryon density from $Planck$.}
\label{fig:cntr_xi_eta}
\end{figure}

Within the framework of general relativity, BBN can be, for example, extended to include
neutrino degeneracy (e.g. Ref.~\cite{Weinberg1972book}).
Degeneracy of electron-neutrinos is described in terms of a parameter
\begin{equation}
\xi^{}_{\rm e} = \mu^{}_{\nu,e}/kT^{}_\nu,
\end{equation}
where $\mu^{}_{\nu,e}$ is the chemical potential of electron  neutrinos
and $T_\nu$ is the temperature of neutrinos.
To get  abundance variations of both neutrons and protons against
$\xi_{\rm e}$, we take a usual method to incorporate
the degeneracy into the Fermi-Dirac distribution of neutrinos~\cite{Weinberg1972book}.
In this study, we do not consider the degeneracy of $\tau$- and $\mu$-neutrinos. 

In BBN calculations, we implemented the neutrino degeneracy as follows.
Before the temperature drops to the difference $Q/k$ in the rest mass energies between
a neutron (n) and 
{{a}}
proton (p), they are in thermal equilibrium through the weak interaction
{{prococesses:}}
$\rm n+e^+ \rightleftharpoons p+\overline{\nu}_e$, $\rm n+\nu_e  \rightleftharpoons p+e^-$, 
{{and}} $\rm n \rightleftharpoons p+e^+ +\overline{\nu}_e$.
Below $T = 4$~MeV, we solve the rate equations for n and p until $T$ drops to 1 MeV including 
{{the individual}} weak
interaction 
{{rates.}} After that,
we begin to operate the nuclear reaction network with the weak
interaction rates between n and p included.
We should note that in the present parameter range shown later,  effects
of neutrino degeneracy on the expansion
and/or cooling of the universe can be almost neglected, because the
absolute values of neutrino degeneracy are
rather small effects on energy density at most $10^{-3}\ \%$. (see Fig.\ref{fig:cntr_xi_eta}).  

The produced amounts of D and ${}^7_{}$Li are almost the same {{compared to the case of
the standard BBN}}, while
${}^4_{}$He becomes less abundant if $\xi^{}_{\rm e} > 0$, because
$\beta$-equilibrium {{leads to lower neutron production.}} 
This is because, the abundance ratio of neutrons to protons (n/p) is
proportional to $\exp [- \xi_{\rm e}]$.
This can be seen in Fig.\ref{fig:bbn_xi_nacre2}; {{while}} the abundance of
 ${}^4_{}$He is very sensitive to $\xi^{}_{\rm e}$,
it is insensitive to $\eta$. On the other hand, although the abundance of D is almost uniquely
determined from $\eta$, i.e., the nucleon density, it depends weakly on
$\xi^{}_{\rm e}$. t

When $\xi^{}_{\rm e}$ increases, the produced amount of ${}^4_{}$He decreases. 
{{On the other hand, when}} 
$\xi^{}_{\rm e}$ becomes negative, more neutrons
survive to yield more $^4$He as seen in Fig.\ref{fig:bbn_xi_nacre2}.
It should be noted 
that {{the production of D is only weakly affected by $\xi^{}_e$.}}

To find reasonable values of $\xi^{}_{e}$ and $\eta^{}_{10}$ which satisfy the consistency between
BBN and observed $^4$He and D, we
calculate $\chi^{2}_{}$ as follows:
\begin{equation}
 \chi^{2}_{}(\eta,\xi^{}_e)=\sum_{i}{\frac{\left( Y^{th}_i(\eta,\xi^{}_{e}) -
			     Y^{obs}_{i}\right)^2}
{{\sigma^{2}_{th,i}}+{\sigma^{2}_{obs,i}}}}, 
\label{eq:chisq}
\end{equation}
where $Y^{}_{i}$ and $\sigma^{}_i$ are the abundances and their uncertainties
for elements $i~(i=Y_p, {\rm D})$, respectively.
The value $\sigma_{th,i}$ is obtained from the Monte-Carlo calculations using 1$\sigma$ errors associated with 
nuclear reaction rates. 
The observational values, $Y^{obs}_i$ and their errors $\sigma^{}_{obs,i}$, are taken from \eqref{eq:Yp_Izotov13} and \eqref{eq:D_Cooke14}. 

Figure~\ref{fig:cntr_xi_eta} shows the contours having 1$\sigma$,
2$\sigma$, and 3$\sigma$ confidence levels~(C.L.) on the
$\eta^{}_{10}-\xi^{}_{\rm e}$ plane obtained \eqref{eq:chisq}.

In consequence, we get the following constraints for both $\eta_{10}$ and
$\xi_{\rm e}$ with the $1\sigma$~C.L.:
\begin{equation}
 6.17 < \eta^{}_{10} < 6.38 \,\,\,\ \
 -3.4\times10^{-2} <\xi_{\rm e} < -1.8\times10^{-2},
\label{eq:xi_eta_results_1}
\end{equation}
and with the $2\sigma$ C.L.:
\begin{equation}
  6.02 < \eta^{}_{10} < 6.54 \,\,\,\
 -4.6\times10^{-2} <\xi_{\rm e} < -0.4\times10^{tb-2}.
\label{eq:xi_eta_results_2}
\end{equation}
It is noted that, except for neutron decay, two-body reactions are
dominant during BBN. The weak reactions are only $\beta$-decay
of ${}^3$H with $\tau^{}_{1/2}=12.33$~y and e-capture of ${}^7$Be
with $\tau^{}_{1/2}=53.29$~d~\footnote{ENSDF~{\it http://www.nndc.bnl.gov/ensdf/index.js}}. These half lives are modified
by a small factor through neutrino degeneracy. However, the
final abundance is not affected at all.



\section{Discussion}
\label{sec:result}
While a large spread in errors of $^4$He by Aver et al.~\cite{Aver2013}
hinders us from constraining the amount of the produced $^4$He abundance,
{{a smaller range by Izotov et al.~\cite{Izotov2013} 
permit us to constrain the ${}^4_{}$He production}}.
Our results clarify the
present controversial situation between standard BBN and observations; the effects
of uncertain mechanism originated from a non-standard theory should reflect the ratio of n/p.

If we adopt the value in \eqref{eq:xi_eta_results_2}, 
{{we can obtain the following range for vdensity parameter:}}
\[
0.0220 \le \Omega^{}_b h^2 \le 0.0239, 
\]
which is compatible with {{that}} from $Planck$ measurements.
{{We showed}} that the neutrino degeneracy may
become one of solutions to solve the discrepancy concerning the present
baryon density between BBN and CMB.
{{Our results provide a}}
{{{narrower range of $\xi^{}_e$ compared with 
{{the}}
previous
{{study,}} e.g.~Ref.\cite{Kneller2001}}. 
BBN alone seems to give 
{{a}} strong constraint on parameters of a non-standard model such as 
the neutrino degeneracy.}

The ${}^7_{}$Li abundance in the present calculation is still larger than
the observational values~\eqref{eq:Li7_Sbord} and \eqref{eq:Li7_Korn}. 
For the apparent discrepancies among the nuclear data and observations,
we may need {{a}} non-standard model beyond Friedmann model:
For example, the expansion rate in the universe could deviate
significantly in a framework of a Brans-Dicke
theory~\cite{Arai1987,Etoh97,Nakamura2006,Berni2010}
{{{, or a scalar-tensor theory of
gravity~\cite{Coc2006rt,Larena2007}}}}.

If inhomogeneous BBN~\cite{Applegate1987,Rauscher1994} could occur in some regions in the universe, it may
solve the problem concerning ${}^7_{}$Li abundance: {{{if
there is 
{{a}} high density region of $\eta > 10^{-5}$ in the BBN era,
amounts of produced ${}^7$Li decreases significantly. As a consequence,  
the average value of ${}^7$Li between the high and the low density
regions becomes lower than the predicted value in SBBN~\cite{NakamuraIBBN2013}}.
}}}

Finally, we would like to emphasize that the nuclear reaction
rates responsible to the production of He and D are still not definite. The error
bars given by NACRE~II~\cite{NACRE2} may not be always confirmed by other experimental groups. 

\begin{acknowledgements}
This work has been supported in part by a Grant-in-Aid for Scientific
Research (24540278) of the Ministry of Education,
Culture, Sports, Science and Technology of Japan.
\end{acknowledgements}

\bibliography{ichimasa_bibfile02}

\end{document}